\documentclass[amsmath,amssymb,aps,twocolumn]{revtex4-2}
\usepackage{graphicx}
\usepackage{subfigure}
\usepackage{bm}

\begin{document}

\title{Quantum Superposition States: Spin Glasses and Entanglement}

\author{Asl{\i} Tuncer}
    \email[Correspondence email address: ]{asozdemir@ku.edu.tr}
    \affiliation{Ko\c{c} University, Institute of Physics, Istanbul,Turkey}
 \author{Serhat C. Kad{\i}o\u{g}lu}   
 \email[Correspondence email address: ]{asozdemir@ku.edu.tr}
    \affiliation{Ko\c{c} University, Institute of Physics, Istanbul,Turkey}

\date{\today} 

\begin{abstract}
Spin-glass (SG) is a fascinating system that has garnered significant attention due to its intriguing properties and implications for various research fields. In condensed matter physics, SGs are a prototypical example of disordered systems and have been studied extensively to understand the behavior of complex systems with the random disorder. One of the key characteristics of SGs is that they contain random disorder, which leads to many possible states of the system occurring with very close probabilities. We explore the concept of spin-glass superposition states (SSs), which are equiprobable SSs of possible electronic configurations. Using the Edward-Anderson (EA) type SG order parameter and magnetization, we demonstrate that these SSs can be classified based on their contribution to distinguishing magnetic order (disorder), such as SG, (anti)ferromagnetic, and paramagnetic phases. We also generalize these SSs based on the system size and investigate the entanglement of these phase-based SSs using the negativity measure. We show that the SG order parameter can be utilized to determine the entanglement of magnetically ordered (disordered) phases, or vice versa, with negativity signifying magnetic order. Our findings provide further insight into the nature of quantum SSs and their relevance to SGs and quantum magnets. They have implications for a range of fields, including condensed matter physics, where SGs are a prototypical example of disordered systems. They are also relevant for other fields, such as neural networks, optimization problems, and information storage, where complex systems with random disorder behavior are greatly interested. Overall, our study provides a deeper understanding of the behavior of spin glasses and the nature of quantum SSs, with potential applications in a variety of fields.
\end{abstract}

\maketitle

\section{Introduction} \label{sec:intro}

Spin glasses are a fascinating phenomenon in condensed matter physics due to their unique microscopic properties. However, these systems contain random disorder, resulting in many possible states occurring with similar probabilities~\cite{FH1991,Mydosh,nish} and being unable to arrange in a particular spin state, which satisfies the energy minimum for each interaction~\cite{Toulouse77II, Harary}. Such a situation is called frustration~\cite{Toulouse77I}. Even if a system is unfrustrated classically, it may exhibit frustration in the quantum case~\cite{illum2011,wolf2003,daw04,giampaola10,beaudrap10} due to non-commutativity and entanglement~\cite{daw04,adhikari2009}. Interestingly, even with just a few entangled elements, novel phenomena can occur in the quantum domain~\cite{daw04,adhikari2009,Parisi,Lewenstein}. Quantum fluctuations and entanglement can also play essential roles in the behavior of spin glasses since the spin glass order occurs at low temperatures~\cite{Parisi}, and thermal fluctuations do not dominate the feature of spin-glass order. Quantum interference may also lead to unexpected effects, such as the suppression of tunneling \cite{Razavy} or the formation of localized states \cite{anderson,MORO2018}. Spin glasses, on the other hand, should have frustrated spin(s). When considered from the quantum perspective, the complex behavior of spin glasses is understandable as arising from quantum interference and entanglement of formation.

The frustration results in a complex and disordered arrangement of the magnetic spins, which can be described by a distribution of spin configurations rather than a well-defined pattern. Since frustration leads to many local minima in their free energy landscapes, making it difficult to choose any configuration due to their equiprobable nature. In contrast to this approach, we investigate the existence of spin glasses in distinct quantum superposition states of possible electronic configurations without needing any ensemble of spin configurations. As a result, it can be thought that spin glasses become frozen in any of these configurations. We directly measured the local magnetizations of the spins and the Edwards-Anderson SG order parameter to describe the corresponding magnetic phases, which include all-to-all interactions and randomly distributed antiferromagnetic (AFM) impurities. The paper is organized in Section~\ref{sec:model} we introduce our model and describe the procedure for identifying the superposition states that contributing to the SG phase. We also expand our results to well-known magnetic orders (disorder) and classify these superposition states (SSs) concerning their phase contributions, such as paramagnetic (PM), ferromagnetic (FM), and antiferromagnetic magnetic phases in Section~\ref{sec:develop1}. Once we identify the phase-based superposition states, we discuss the role of entanglement and the relationship between the SG-order parameter and entanglement in Section~\ref{sec:entnglmnt}. Finally, we conclude the paper in Section~\ref{sec:conclusions} with an outlook of our results and the impact on future theoretical investigations and experimental implementations of current quantum technologies.

\section{Model}
\label{sec:model}
We consider $N$ Ising spins interacting through infinite-ranged exchange interaction with the Hamiltonian,
\begin{equation}
    H=-\Sigma_{(i,j)} J_{ij} \sigma_i^z\sigma_j^z.
    \label{eq1:HEA}  
\end{equation}
Here $\sigma_i^z$ is Pauli z-matrices with $i, j=1,\dots,N$ and the interaction couplings are quenched variables governed by a Gaussian distribution with a variance $J^2/N$ and zero mean $<J_{ij}>=0$, $P(J_{ij})\propto \exp {\frac{1}{2}(\frac{NJ_{ij}^2}{J^2}))}$. The randomly distributed antiferromagnetic interaction is the source of frustration in case the system's Hamiltonian can not be minimized, at least for one spin or bound. Although there is no direct analogy between the geometric frustration in classical systems and its quantum counterpart, it has been defined in quantum systems related to entanglement and coherence effects~\cite{illum}. 
In this study, we will concentrate mainly on $N$-atom system with all-to-all interactions. Since each spin has two states, these Ising spin systems exhibit an exponentially large phase space of $2^N$ configurations. 
 
\begin{figure}
    \centering
    \includegraphics[scale=.06]{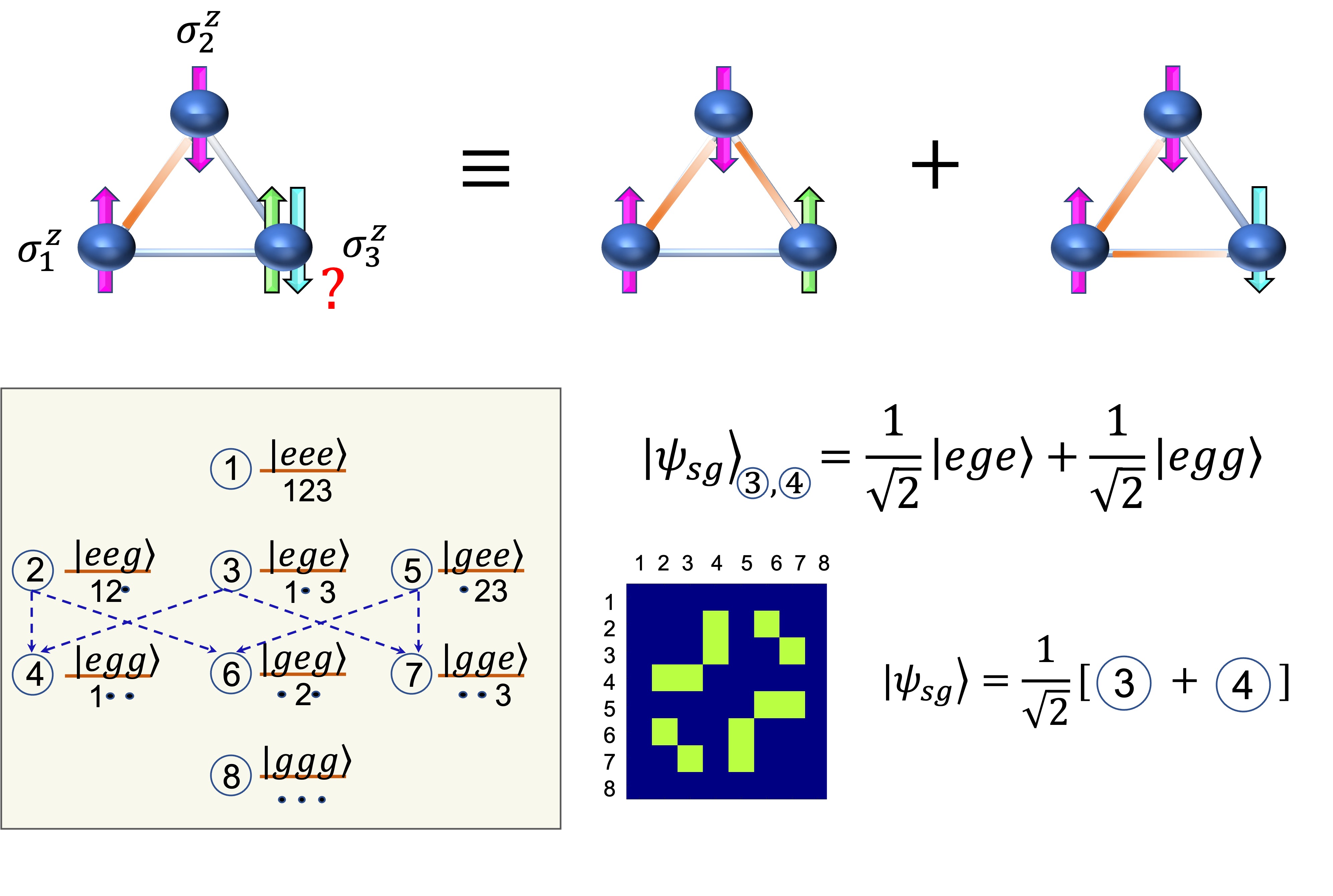}
    \caption{{\it Top panel}: The graphical representation of SG-superposition state for $N=3$ qubits. The interactions (lines) between the qubits (blue spheres) are illustrated for FM $(J>0)$ and AFM $(J<0)$ couplings are yellow and light blue edges, respectively. {\it Bottom left}: The possible SS is found by matching the dashed lines. {\it Bottom right}: One of the $|\psi_{SG}\rangle$ state contributes to the SG-order. The numbers below the levels correspond to the labels of the excited atoms of the state, and dots represent the placement of the ground-level atoms. The circled numbers next to the levels show the corresponding computational basis states. In the smaller box, the corresponding SG-contributed superposition states of the three-spin system with vanishing magnetization and non-zero SG-order parameters are shown in green color.}
    \label{fig:1}
\end{figure}

We started from the simplest model with interacting three-qubit system may have frustration, see Figure~\ref{fig:1}. In the classical case, to see the phase transition, the system should go to the thermodynamic limit, so it would be impossible to see the phase transition on three-spin model. In addition, the thermal or quantum fluctuations drive the system to transition, but we do not consider the thermal fluctuations in this work. 

To inject the quantum effects into the classical version of the Ising model so-called Edward-Anderson spin-glass Hamiltonian~(\ref{eq1:HEA}), one of the well-known ways is to add a transverse field. The quantum fluctuations arises from a competition between the spin-spin interactions and such an applied external field. In contrast to this approach, the present study assumes that quantum fluctuations and frustration are introduced into the system via direct injection of quantum superposition states. Through measurement of corresponding order parameters, such as the local magnetization of the $i^{th}$ spin for a given realization $\alpha$, with $m_i^\alpha= <\sigma_i>_\alpha$, we were able to determine the specific superposition states that contribute to various magnetic phases. The EA spin-glass order parameter, which corresponds to overlaps of the local magnetization~\cite{ea75} and is provided in (\ref{eq2:qEA}), was also utilized in our analysis.
\begin{equation}
    q_{EA}^\alpha=\frac{1}{N}\sum_{i=1}^N (m_i^\alpha)^2.
    \label{eq2:qEA}  
\end{equation}     
As we modified these definitions to our SS concept by taking the averages over our superposition states, the average magnetization is
\begin{equation}
    m=\frac{1}{N}\sum_i^N \langle \psi_{suppos.}|\sigma_i^z|\psi_{suppos.}\rangle,
    \label{eq:msuppos}
\end{equation}
and the spin-glass order parameter includes the overlapping between the discrete states of the entities of the SS by definition. In this way, we have found some contributions to SG order from the equally weighted superposition state in both energy and computational basis with the non-zero $q_{EA}$ and $m=0$. Besides obtaining the SSs contribute to the SG phase, we obtained the PM and (anti)ferromagnetic SSs in both vanishing $q_{EA}=0$ and $m=0$, and both are non-zero cases, respectively. All these different order and disorder SSs are given in Section~\ref{sec:develop1} and we explained that explicit classification and association with their entanglement in Sections~\ref{sec:entnglmnt}.

The state of a system consisting of $N$ spins can be considered as a product state, and the initial state is set to be a superposition of all possible configurations, $|\psi\rangle=\otimes_i^N |\rightarrow\rangle$ with $|\rightarrow\rangle=\frac{1}{\sqrt{2}}(|\uparrow\rangle+|\downarrow\rangle)$, where $|\uparrow\rangle$ and $|\downarrow\rangle$ are the eigenbasis of the Pauli-$z$ operator. Each spin has two possible orientations, up or down, in this representation. However, the superposition states that we have created by taking the sum of these product states will no longer be product states with the quantum correlations they contain. While the random disorder interactions between spins force a fixed orientation to minimize the energy, some of the spins may remain in the $|\rightarrow\rangle$ state even if there is no external field due to frustration.

In Figure~\ref{fig:1}, the antiferromagnetic $(J<0)$ interactions may cause geometric frustration in the first triangle, as shown by the two-side-aligned arrows denoting the frustrated spin. The qubit state can be in any superposition of the form  $|\psi\rangle=a_1|0\rangle+a_2|1\rangle$ as long as $|a_1|^2+|a_2|^2=1$. However, this is not the case for classical spins. We illustrate such a frustrated configuration state of a three-body system separated into two states that do not have frustration in the $\sigma^z$ basis:
\begin{equation}
\begin{aligned}  |\psi>&=|0\rangle\otimes|1\rangle\otimes\left (\frac{|0\rangle+|1\rangle}{\sqrt{2}}\right ),\\
    &=\frac{1}{\sqrt{2}}\left (|0\rangle\otimes|1\rangle\otimes|0\rangle+|0\rangle\otimes|1\rangle\otimes|1\rangle \right ).
\end{aligned}
\label{eq:4}  
\end{equation}  
The corresponding phases can be obtained from the relative order parameters. Therefore, although many different superposition states could be considered via the energy eigen states or computational basis vectors, we will only consider the equally weighted two-state superposition in the computational basis. 
We will continue with the standard basis of energy-state products from now on. This standard basis is ${|e\rangle, |g\rangle}$ for one atom, ${|ee\rangle, |eg\rangle, |ge\rangle, |gg\rangle}$ for two, and ${|eee\rangle, |eeg\rangle, |ege\rangle, |gee\rangle, |egg\rangle, |geg\rangle, |gge\rangle, |ggg\rangle}$ for three atoms. Figure~\ref{fig:1} shows the three-atom standard basis on the left bottom. Corresponding positions in the natural basis of the levels are given in the circles next to levels and the numbers with the green dots below the levels denotes the placement of the excited atoms and ground state, respectively. 

All our superposition states contributes to the spin-glass order should satisfy the two simple rules:
\begin{itemize}
    \item[(i)] The SG-contibuted SSs must have at least one co-excited spin, 
    \item[(ii)] The total number of excited spin labels should not exceed the total number of spins in the system.
\end{itemize}

The first rule denotes the overlapping between the states, and the second one corresponds to the SS should have at least one frustrated spin an equally-weighted qubit-state such as called \textit{cat state} denoted by \textbf{C}. Once we address the excited spins, we can write the spin-glass SSs quickly in each system size. However, in ferromagnetic (antiferromagnetic) SSs, the number of labels can be greater than the number of spins in the system. Understandably, the ferromagnetic order does not have to include any frustrated spin, and all the spins can be in the same (up or down) direction. It can be classified as that one excited and a set of cat states \{\textbf{C},\textbf{C},\dots \} pair will contribute to the FM/AFM orders. To fix a spin on a state can be thought of as making a local measurement on it, and this causes a loss of quantumness. We examined this loss in terms of negativity, a measure of quantum entanglement defined as~\cite{Vidal},
\begin{equation} 
\mathcal{N}=\frac{\lVert \rho^\mathrm{T} \rVert_1-1}{2}. 
\label{eq:5}
\end{equation}
Here, the trace norm of the quantum state $\rho=|\psi_{suppos}\rangle \langle\psi_{suppos}|$ is denoted by $\lVert \rho^\mathrm{T} \rVert_1$. The detailed about negativity calculations and results are explained in Section~\ref{sec:entnglmnt}.

\section{Quantum Superposition States for Magnetic Orders and Disorder} \label{sec:develop1}
The magnetic phases of interest can be classified in terms of magnetization and the EA spin-glass order parameter,  $q_{EA}$~\cite{ea75}.
\begin{figure}[ht]
\centering   
    \subfigure[Order parameters for equally weighted SSs.]{
    \includegraphics[width=.9\linewidth]{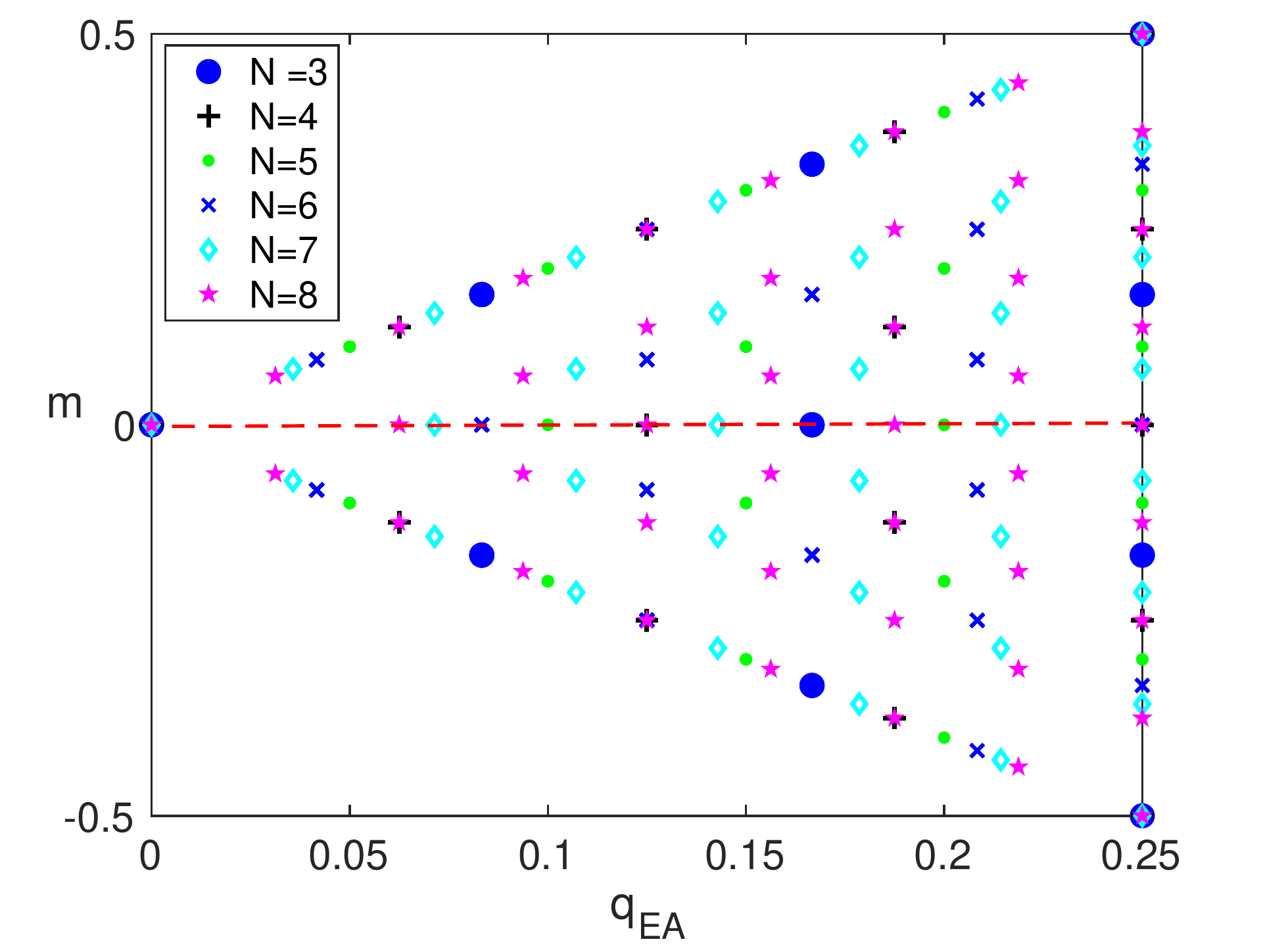}   
    \label{fig:2a}
    }

    \subfigure[Order parameter for all superposition states.]{
    \includegraphics[width=.9\linewidth]{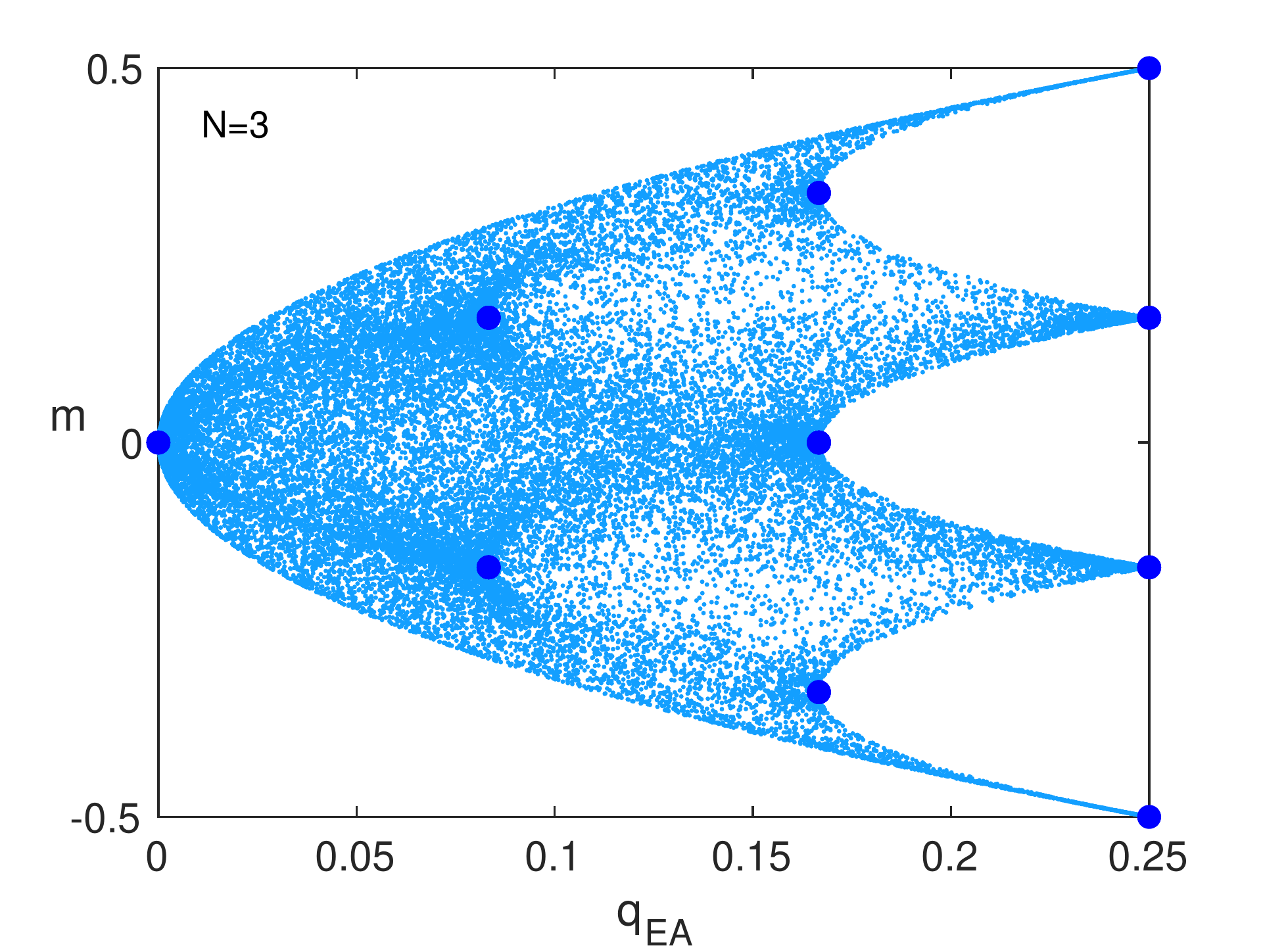}
    \label{fig:2b}
    }
\caption{(Color online.) The magnetization vs. EA spin-glass order parameter, $q_{EA}$, is shown for (a) all equally-weighted superposition states in different sizes $N=3,4,5,6,7,$ and $8$. The red horizontal line indicates $q_{EA}\neq 0$ with $m=0$, corresponds to the spin-glass regime, (b) not only the equally weighted SSs but also expanded superposition states at $N=3$. We obtained expanded superposition states by giving different weights to superposed ones. In both figures, the paramagnetic phase is observed at the point $m=0$ and $q_{EA}=0$. These findings provide valuable insights into the nature of quantum SSs and their relevance to SG and quantum magnets.}
\label{fig:2}
\end{figure} 
All corresponding magnetization and $q_{EA}$ parameter values of the equally weighted and expanded superposition state spaces are displayed in Figure~\ref{fig:2a} and Figure~\ref{fig:2b}, respectively. The parameter values of equally weighted superposition states, a special case of the expanded ones, are obtained at N=3,4,5,6 system sizes. 
\begin{figure*}[ht]
    \includegraphics[scale=.28]{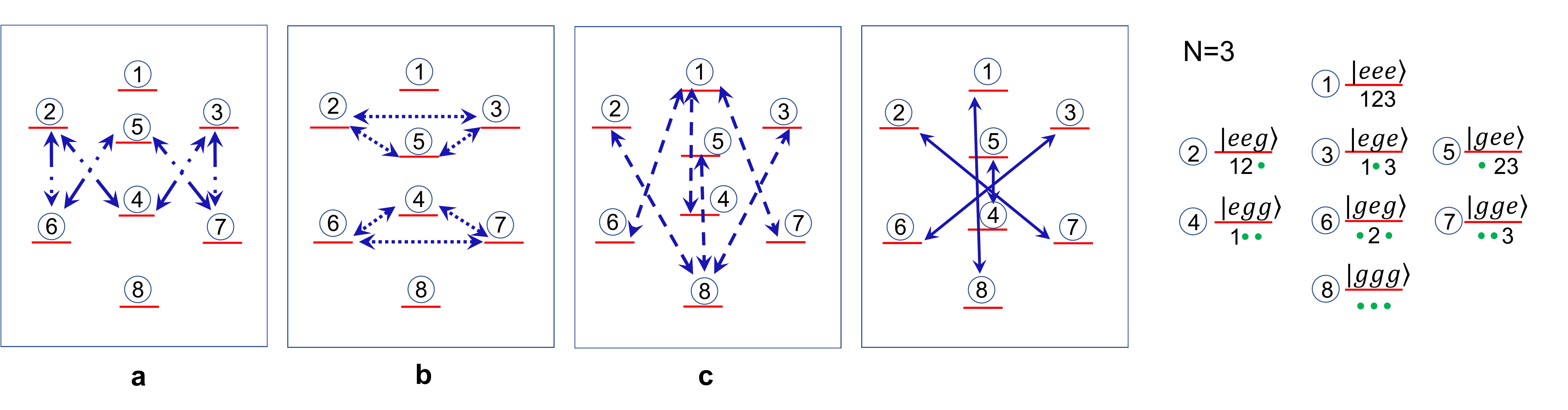}
    \caption{(Color online.) Schematic representation of SSs contributes to (a)SG, (b)FM, (c)AFM and (d)PM phases. Arrows depict the equally weighted summation of the states. This figure provides a clear and comprehensive visualization of the various superposition states and their relation to spin glasses, quantum magnets, and quantum superposition states.}
    \label{fig:3}
\end{figure*} 

The triangular parameter domain can be seen in all cases; moreover, the maximum value of the $q_{EA}$ remains the same. In each system size, it is possible to see all the magnetic phases in interest. 

In both figures, having various numbers of $m$ values against the same $q_{EA}$ value indicates spontaneous symmetry breaking~\cite{spontsymBreak}. While this symmetry breaking signifies the FM (AFM) order, the PM phase corresponds to the point of $m=0$ and $q_{EA}=0$. The red-dashed line starts with the PM regime, and along the line, we can obtain the SSs corresponding to the SG regime. For simplicity, we will continue with the particular case of these SS space, such as equally weighted superposition states. However, even if the system size enlarges and extra SSs arise, the triangular structure of the diagram remains the same. In other words, the symmetry breaking can be observed for $N\to \infty$ in terms of the different $q_{EA}$ values.

\begin{figure*}
\centering   
    \subfigure[The non-zero SG order parameter $q_{EA}$ of equally weighted SSs.]{
    \includegraphics[width=.98\linewidth]{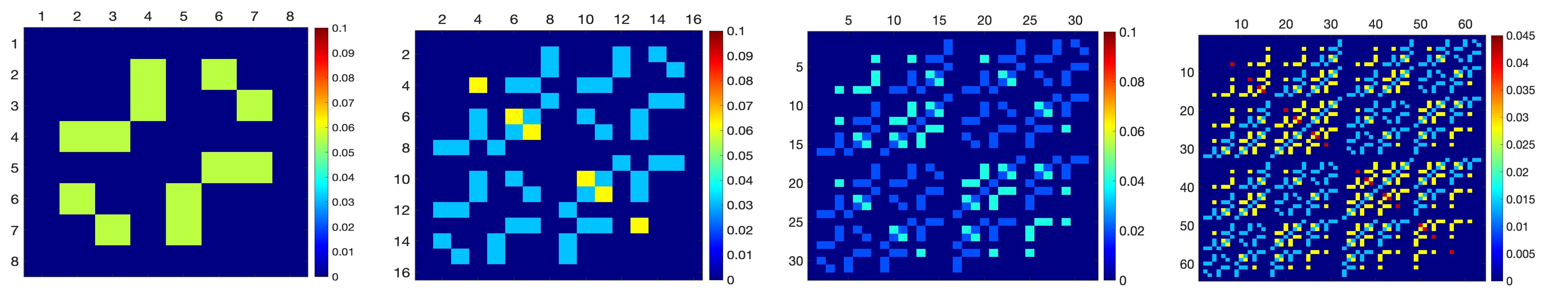}   
    \label{fig:4a}
    }

    \subfigure[First line: The Entanglement of equally weighted SSs contribute to SG order. Second line: Entanglement of all magnetic orders. ]{
    \includegraphics[width=.98\linewidth]{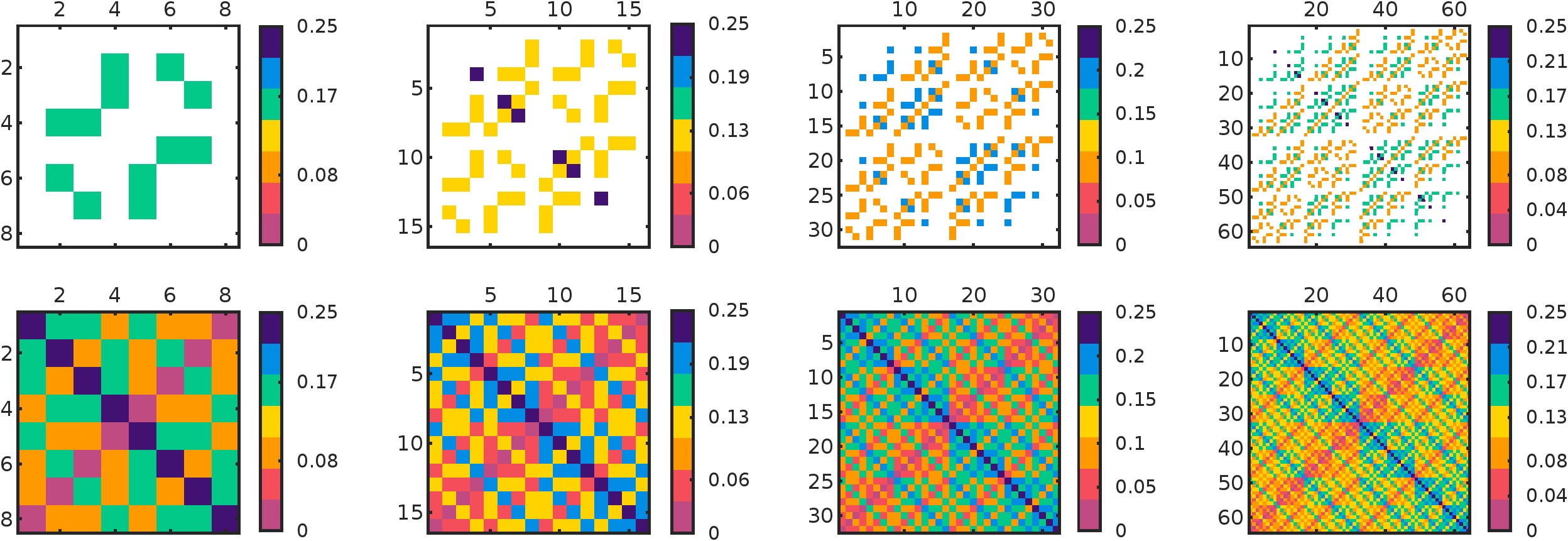}
    \label{fig:4b}
    }
\caption{(Color online.) (a) The matrix representation of all spin-glass order  contributed equally-weighted superposition states is given for $N=3,4,5$, and $6$ spins from left to right with non-zero Edward-Anderson order parameter and zero magnetization. As the system size increases, the number of distinct values of $q_{EA}$ increases, indicating a signal of replica symmetry breaking. The various colors denote different values of $q_{EA}$ (or $\mathcal{N}$), and this pattern repeats itself in subsequent generations from left to right. Diagonal elements correspond to the single states and they are equal to $q_{EA}=0.25$ fixed value. However, they contribution to spin-glass or FM/AFM order may alter by the system size. One another fixed part of the matrix representation is off diagonal elements have only $q_{EA}=0$ values and they contributes to the PM order for all system sizes. (b) The same pattern of $q_{EA}$ can be obtained for average negativity $\mathcal{N}$ with different numerical values. It should be noted that there is a reciprocal relationship between the $\mathcal{N}$ average negativity and the $q_{EA}$ order parameter.}
\label{fig:4}
\end{figure*}

Let's return to the SSs that are equally weighted. Figure~\ref{fig:3} shows how binary superpositions in a system of dimension $N=3$ correspond to different quantum magnets, including SG, FM, AFM, and PM-phases. As shown in Figure~\ref{fig:4a}, we have derived the matrix representation of the binary states that contribute exclusively to the SG phase in systems comprising $N=3,4,5,$ and $6$ atoms. In order to maintain clarity, we have elected to exclusively represent the SG-SSs, omitting the explicit depiction of the superposition states that would pertain to other magnetic phases. However, it is pertinent to note that the off-diagonal elements do indeed play a role in the PM phase, as they are composed of non-overlapping product spin states. Furthermore, the upper and lower triangular regions of the non-diagonal elements in the matrix are associated with the FM or AFM phase, depending on the specific binary superpositions involved. Consequently, this superposition matrix can be regarded as an evenly distributed superposition state space that pertains to the configuration space. The matrix is divided into distinct quantum magnetic phases, thereby serving as a phase diagram. Additionally, we have noted that the phase partition pattern remains consistent and scalable even as the system size is expanded.
The differentiation of the $q_{EA}$ order parameter at each even value of system size, $N$, can be associated with replica symmetry breaking in spin-glass systems~\cite{Parisi}. For example, while there is a unique $q_{EA}$ value for $N=3$, sizes $N=4$ and $N=5$ have two different values of $q_{EA}$. Figure~\ref{fig:4} illustrates a recursive pattern, wherein taking a partial trace over the most recently added qubit leads the system state space to the previous state space. Co-excited atoms in SSs can also be defined as overlapping between two states. This overlap scales with the system size, similar to the differentiation of $q_{EA}$. The number of possible overlapped atoms increases with each even number of $N$. However, unlike the SG-SSs, the PM-phase SSs have neither co-excited spins nor any overlap. All PM-phase SSs are maximally entangled, similar to Greenberger-Horne-Zeilinger (GHZ) states~\cite{ghz}. We illustrated that entanglement of spin-glass SSs in the first line of the Figure~\ref{fig:4b} and entanglement of SSs corresponds to the other magnetic phases in the second line of the Figure~\ref{fig:4b}, respectively.

\section{The New Entanglement Witness of the Magnetic Structures} \label{sec:entnglmnt}
\begin{figure*}[ht]
\centering
\includegraphics[scale=.4]{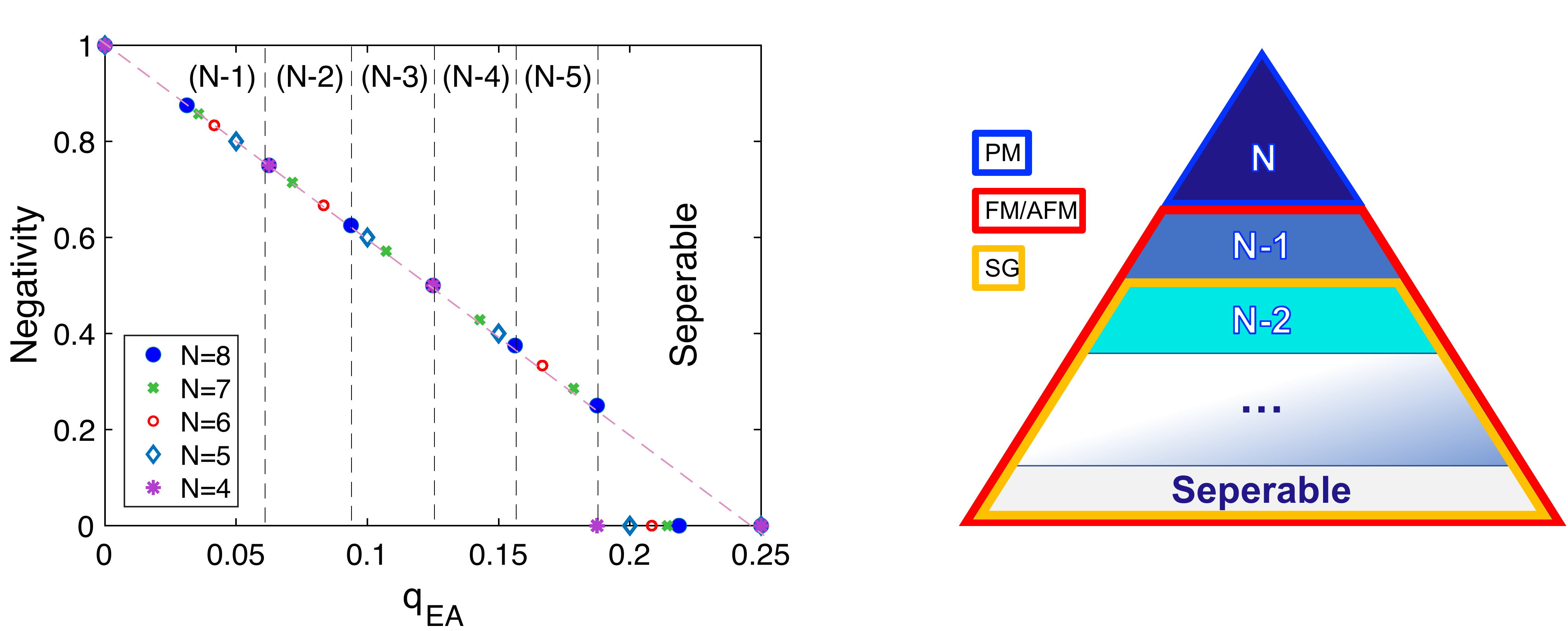}
\caption{(Color online.) The left panel shows the variation of the negativity with the Edward-Anderson spin-glass order parameter $q_{EA}$ for particle systems with $N=4,5,6,7$, and $8$. The distinct regions in the plot correspond to different values of $q_{EA}$ that indicate the extent of particle entanglement. Initially, all systems are fully entangled with $\mathcal{N}=1$, and as $q_{EA}$ decreases, the number of entangled particles decreases, ultimately leading to a separable state with $\mathcal{N}=0$. In the right panel, the phase-contributed superposition states are classified based on the number of entangled particles in the $N$-particle system.}
\label{fig:5}
\end{figure*}

The presence or absence of overlap between states in their superposed states corresponds to the ordered/disordered states of the system. Moreover, the overlapping superposition states can exhibit different magnetic orders, such as spin-glass, ferromagnetic, and antiferromagnetic orders. Firstly, after defining the order/disorder distinction based on the presence or absence of overlap, we observe that paramagnetic (disordered) systems corresponding zero overlaps also possess maximum entanglement. From this perspective, we calculated the entanglement of the superposition states using logarithmic negativity to define the relationship between the amount of overlap and entanglement. Considering that the Edward-Anderson spin glass order parameter measures the degree of overlap between two different system configurations (superposed states), we investigated the direct relationship between negativity $\mathcal{N}$ and $q_{EA}$ order parameter. The numerical results illustrate the relationship between these two quantities, as shown in Figure~\ref{fig:5}.
We obtained that the order parameter $q_{EA}$ decreases linearly with $\mathcal{N}$,
\begin{equation}
q_{EA}=q_{max}-\frac{1}{4}\mathcal{N}.
\label{eq:6}
\end{equation}

Here $q_{max}=1$ is the maximum value of the normalized EA spin-glass order parameter. According to the figure, the state starts from the fully entangled state (PM state) with $\mathcal{N}=1$ and $q_{EA}=0$. Then, its maximal-entangled portion becomes smaller and smaller until it reaches the separable state with $\mathcal{N}=0$ and $q_{EA}=q_{max}=0.25$. Each distinct $q_{EA}$ value corresponds to a different entangled cluster size. We separated regions corresponding to $m$-partite entanglement with $m=N,N-1,\dots,1$ by dashed lines. However, as the system size approaches infinity ($N\to \infty$), the classification mentioned above is no longer discernible, and the different $q_{EA}$ values on the fitting line be indistinguishable. This is due to the loss of quantumness in the system, as it becomes classical and the separations between distinct $q_{EA}$ values vanish.

In this analysis, we present both numerical results and a discussion of superposition states, including the multi-partite entangled component, concerning the recursive growing pattern of the sg-order parameter $q_{EA}$ and negativity $\mathcal{N}$, illustrated in Figure~\ref{fig:4}. Specifically, we consider a three-particle system where each particle can exist in one of the three configurations from the set $E_{N=3}(SG)=\{\textbf{C},\textbf{e},\textbf{g}\}$, where $E_{N=3}(SG)$ denotes the ensemble including the probable spin configurations. We identify six possible configuration states for the SG contribution with vanishing magnetization and non-zero $q_{EA}$. 

Suppose one more qubit is added to the system. The system state has two possible configuration ensembles: 
\begin{equation}
E_{N=4}(SG) \\ 
=
\begin{cases}
   \{\textbf{C},\textbf{C},\textbf{e},\textbf{g}\}, &q_{EA}=0.125, \\
   \{\textbf{e},\textbf{g},\textbf{e},\textbf{g}\}, &q_{EA}=0.25. 
\end{cases}
\label{eq:7}
\end{equation}
which correspond to SG-contributed superposition. 
In this case, we observe two distinct values of $q_{EA}$ corresponding to the two different configuration sets. The permutation group of the first set yields 24 different SSs that have $q_{EA}=0.125$, while the second set yields six SSs with $q_{EA}=0.25$ appropriate to Figure~\ref{fig:4}. As we increase the system size to $N=5$ spins, the number of possible sets remains the same as for $N=4$, and we observe two additional sets, namely $\{\textbf{C},\textbf{C},\textbf{C},\textbf{e},\textbf{g}\}$ and $\{\textbf{C},\textbf{e},\textbf{g},\textbf{e},\textbf{g}\}$. 

Notably, the $q_{EA}$ parameter differs between distinct sets and within a set, depending on the number of entangled particles in a superposition state. For instance, the set $\{\textbf{C},\textbf{C},\textbf{C},\textbf{e},\textbf{g}\}$ can be considered as 
\begin{equation}
\{\textbf{C},\textbf{C},\textbf{C},\textbf{e},\textbf{g}\} \\ 
=
\begin{cases}
   \{|GHZ\rangle_3, \textbf{e}, \textbf{g} \} \\
   \{|GHZ\rangle_2,\textbf{C},\textbf{e},\textbf{g}\},  
\end{cases}
\label{eq:8}
\end{equation}
here, the sub set $|GHZ\rangle_3:\{\textbf{C},\textbf{C},\textbf{C}\}$ and $|GHZ\rangle_2:\{\textbf{C},\textbf{C}\}$ gives the maximally entangled (GHZ) states, and the subscript denotes the number of entangled particles. In general definition, $n$-particle GHZ state can be written as 
\begin{equation}
 |GHZ\rangle_n=\alpha(|g\rangle^{\otimes n}+|e\rangle^{\otimes n}),
\label{eq:9}
\end{equation}
where $\alpha$ is a constant~\cite{ghz,nielsen}, the source of these maximally entangled states is related to the number of cat state $C$ in the permutation sets.

If the possible configuration sets lack \textit{cat state} \textbf{C}, the resulting state will be deemed \textit{separable}. The superposition state may also be partially entangled, in which the number of cat states is less than $N-1$. Defining the entangled portion of the state, particularly for larger systems, presents a challenge. While our focus centers on SG superposition states, analogous conditions arise in FM and AFM cases. 

We developed a metric to quantify the entanglement partition of a state in terms of the spin-glass order parameter, $q_{EA}$, which allowed us to achieve our objective. Figure~\ref{fig:5} illustrates the entanglement partitions for various system sizes, along with the corresponding inverse linear relationship between the negativity and $q_{EA}$. As $q_{EA}$ decreases from its maximum value corresponding to separable states, the number of entangled particles increases until the system reaches a state of maximum entanglement. 

Based on the presence of entangled particle ensembles, Figure~\ref{fig:5} provides a classification of magnetic phases. 

\section{Conclusions} \label{sec:conclusions}

This research demonstrates that spin glasses can exist in equiprobable superposition states of potential electronic configurations in a quantum framework. We propose using cat states to define frustrated spins and link the frustration to quantum interference. By employing the Edward-Anderson spin-glass order parameter and magnetization, we classify the superposition states based on their contribution to distinguishing magnetic order (or disorder) in various phases, such as SG, (anti)FM, and PM. Our results provide valuable insights into the nature of spin glasses in quantum systems and have implications for developing quantum technologies such as quantum cryptography~\cite{Pirandola}, quantum simulation~\cite{georgescu,daley2022,brown2010}, quantum computation~\cite{Bennett, Valiev,Kim,Normand}, quantum sensing~\cite{qusens} and metrology~\cite{Schaetz}.

 We establish a direct correlation between the Edward-Anderson spin glass order parameter, $q_{EA}$, and a measure of entanglement, negativity represented by $\mathcal{N}$. We demonstrate that the spin glass order parameter can also function as an indicator of entanglement, while conversely, the negativity of entanglement can serve as the order parameter to distinguish between phases of order and disorder, specifically the ferromagnetic (FM) and paramagnetic (PM) phases. This result is due to the fact that entanglement is the ability of qubits to correlate their state with other qubits.
 Quantum phase transitions (QPTs) are an established finding in condensed matter physics, characterized by significant changes in the ground state properties of a quantum system induced by small variations in an external parameter. For example, QPTs can be induced in spin systems by variations in the magnetic field~\cite{Sachdev, qptscience}, while in cold atom simulators of Hubbard-like models, changes in the intensity of a laser beam can trigger QPTs~\cite{Lewenstein}.
 While our study does not address QPTs directly, several methods for driving quantum states to a target state have been studied extensively in the literature. These include quantum entanglement, state transfer~\cite{stransf,stransf0,stransf1,stransf2,stransf3,stransf4}, and quantum adiabatic evolution~\cite{Joye1994,stransf5}. Once we obtain the corresponding quantum states to the different quantum phases, phase transition can be studied in this contextuality.

This study reveals that the structural similarities between entanglement and spin glass order parameter persist across systems of varying sizes, as indicated by the consistent patterns shown in Figure~\ref{fig:4}. Furthermore, our study highlights the potential use of superposition states in defining magnetic order (disorder)~\cite{Schaetz} in condensed matter physics, which has broader implications for quantum information processing and quantum computing~\cite{Bennett, Valiev,Kim,Normand}. These findings offer new insights into the nature of quantum superposition states and their relevance to spin glasses and quantum magnets. The categorization of states according to their magnetic properties, utilizing physical order parameters and entanglement, is a critical prerequisite for the effective manipulation of entangled states that are necessary for quantum information processing and transfer via qubits~\cite{Bennett1993}. 
We offer that these superposition states are candidates to be new phase-based-bits in usage of the quantum computing~\cite{nielsen,Feynman}. We are currently investigating their possible use in other physical systems.

\section*{Acknowledgements} \label{sec:acknowledgements}
 The authors would like to acknowledge the financial support from the Scientific and Technological Research Council of T\"{u}rkiye (T\"{U}B{\. I}TAK), grant No. 120F100. We would also like to express our gratitude to Ö. E. Müstecapl{\i}o\u{g}lu and M. Paternostro for their insightful discussions.

\bibliography{QSPSG_Reference}

\end{document}